\documentclass[prd,showpacs,twocolumn,superscriptaddress,floatfix,nofootinbib,10pt]{revtex4-2}
\usepackage{setspace}
\usepackage[utf8]{inputenc}
\usepackage{float}
\usepackage{amsmath,amssymb,amsfonts,bm}
\usepackage{graphicx}
\usepackage{multirow}
\usepackage[usenames,dvipsnames]{color}
\allowdisplaybreaks
\usepackage[
colorlinks=true,
linkcolor=blue,
breaklinks=true,
urlcolor=blue,
citecolor=blue]{hyperref}
\usepackage{orcidlink}
\usepackage{subfigure}
\usepackage{soul}
\usepackage{xcolor}
\usepackage{physics}
\usepackage{booktabs}
\usepackage{ulem}
\usepackage{tikz-feynman}
\usepackage{makecell}
\definecolor{green}{rgb}{0,0.6,0}
\newcommand{\mev}{\textrm{ MeV}}

\newcommand{\GXNU}{\affiliation{Department of Physics, Guangxi Normal University, Guilin 541004, China}}

\newcommand{\GXZD}{\affiliation{Guangxi Key Laboratory of Nuclear Physics and Technology, Guangxi Normal University, Guilin 541004, China}}

\newcommand{\IFIC}{\affiliation{Departamento de F\'{\i}sica Te\'orica and IFIC, Centro Mixto Universidad de
		Valencia-CSIC Institutos de Investigaci\'on de Paterna, Apartado 22085,
		46071 Valencia, Spain}}

\begin{document}
	\title{The $K T_{cc}(3875)$ Interaction: Bound State, Scattering Observables and Correlation Function}
	
	\begin{abstract}
	We investigate the interaction of a kaon with the $T_{cc}(3875)$ state, evaluating the scattering amplitude and the correlation function of the system within a molecular picture in which the $T_{cc}(3875)$ is regarded as a $D^* D$ bound state. Since the $KD^*$ and $KD$ interactions dynamically generate the $D_{s1}(2460)$ and $D_{s0}^*(2317)$ resonances, respectively, both of which are significantly bound with respect to their corresponding particle thresholds, the $KT_{cc}(3875)$ interaction is found to be strongly attractive, leading to the formation of two bound states with binding energies of about 63 MeV and 51 MeV, respectively, with respect to the $KT_{cc}(3875)$ threshold. We also determine the scattering observables at threshold, including the scattering length and effective range, and calculate the correlation function. The latter exhibits features similar to those found in previous studies of the scattering of an external particle off a dynamically generated resonance of molecular nature.
	\end{abstract}
	
	\author{Wen-Hao Jia\orcidlink{0009-0001-1170-3540}}
	\GXNU%
	
	\author{Dao-Cong Lin\orcidlink{0009-0009-4064-7275}}
	\GXNU%

	\author{Chao Huang}
	\GXNU%
	
	\author{Wei-Hong Liang\orcidlink{0000-0001-5847-2498}}%
	\email{liangwh@gxnu.edu.cn}
	\GXNU%
	\GXZD%

	\author{Eulogio Oset\orcidlink{0000-0002-4462-7919}}%
	%\author{Eulogio Oset}%
	\email{Oset@ific.uv.es}
	\GXNU%
	\IFIC%

\maketitle

\section{Introduction}\label{sec:Intr}
Atomic nuclei, with a large amount of protons and neutrons, make up stable systems with many baryons. 
The strong interaction between nucleons and the baryon-number conservation are responsible for the stability of such nuclei. 
The question arises as to why we do not have the equivalent systems formed from aggregates of mesons. 
Certainly, one cannot blame the interaction between mesons, since this interaction is usually stronger than that of nucleons (see, for instance, the strong interaction of vector mesons, in particular coupling to total spin $J=2$ \cite{Molina:2008jw,Geng:2008gx,Du:2018gyn}). 
One important difference is that we do not have meson number conservation, hence, large aggregates of mesons can decay to systems with a smaller number of mesons and the large phase space for decay renders the width very large, putting an end to spectroscopy. 
Yet, there can be exceptions, in particular when all the quarks involved in the system have different flavor, since flavor is conserved in strong interactions \cite{Wu:2022ftm}. 
Another stabilizing factor is large spin, since a large angular momentum gives rise to a centrifugal barrier that suppresses the decay; strong binding energy can also hinder the decay. 
In this sense, it becomes less surprising that in Ref.~\cite{Roca:2010tf} several multi-$\rho$ states were formed theoretically which could be associated in mass and width to the existing mesons $f_2(1270)$, $\rho_3(1690)$, $f_4(2050)$, $\rho_5(2350)$ and $f_6(2510)$. 
Similarly, in Ref.~\cite{Yamagata-Sekihara:2010muv} several states were also produced theoretically from a $K^*$ and multi-$\rho$ mesons, in good agreement with mass and width with the $K^*_2(1430)$, $K^*_3(1780)$, $K^*_4(2045)$, $K^*_5(2380)$ meson states. 
As the number of mesons involved in the system increases, so does the width of the states, putting an end to the experimental observability, which is given by the $f_6(2510)$ multi-$\rho$ meson, and the $K^*_5(2380)$ $K^*$ multi-$\rho$ meson. 
Review papers on the systems made of meson aggregates can be seen in Refs.~\cite{Wu:2022ftm,MartinezTorres:2020hus}, where a large amount of theoretical and experimental papers are reported.

Simultaneously, a new window of opportunity has been opened with the experiments on correlation functions, where interactions of particles not accessible in scattering experiments can be studied \cite{Fabbietti:2020bfg,Liu:2024uxn}. 
In particular, of relevance to the present work, the study of correlation functions of stable particles with resonances which, by themselves are molecular states of mesons, promise to enlarge our understanding of multimeson states. 

A breakthrough step in this direction was given by the measurement of the correlation function for the proton interaction with the $f_1(1285)$ resonance \cite{Lauraser}. 
Anticipating that measurement, a theoretical prediction was made in Ref.~\cite{Encarnacion:2026zas} for the $p f_1(1285)$ correlation function, which was confirmed later by the ALICE experiment of Ref.~\cite{Lauraser}.
The theoretical approach relied on the Fixed Center Approximation (FCA) method to deal with the three-body interaction, appropriate to the case in which the $f_1(1285)$ is considered the cluster and the proton the external particle interacting with the cluster. 
The cluster assumed the $f_1(1285)$ to be a molecular state of $K^* \bar K$ \cite{Lutz:2003fm,Roca:2005nm,Garcia-Recio:2010enl,Zhou:2014ila,Geng:2015yta,Lu:2016nlp} (see the long list of tests for this picture reported in Ref.~\cite{Encarnacion:2025lyf}). 
The FCA used in Ref.~\cite{Encarnacion:2026zas} incorporated the unitarization of the $p f_1(1285)$ amplitude, which was formally implemented in Ref.~\cite{Ikeno:2025bsx}. 
There, the ordinary FCA was taken as an optical potential, which was then used within the Lippmann-Schwinger equation to generate the unitary $p f_1(1285)$ amplitude, following the standard path in the study of particle-nucleus scattering.

The study in Ref.~\cite{Encarnacion:2026zas} produced the $p f_1(1285)$ correlation function, but at the same time it predicted the existence of a bound state of the system with about $35 \mev$ binding. 
This was a new prediction of a three-meson bound state, to be added to the long list of predictions already made (see review paper \cite{MartinezTorres:2020hus}).

Encouraged by the success of the predictions in Ref.~\cite{Encarnacion:2026zas}, and the likely continuation of experiments of this kind by the ALICE collaboration, a wave of calculations followed to make predictions for analogous systems: 
the $n \bar D^*_{s0}(2317)$ scattering was studied in Ref.~\cite{Ikeno:2025bsx}, the $n D_{s1}(2460)$ and $n D_{s1}(2536)$ scattering in Ref.~\cite{Agatao:2025ckp}, the $K f_1(1285)$ in Ref.~\cite{Jia:2026dpl}, the $K D^*_{s0}(2317)$ in Ref.~\cite{Jia:2026iqo}, the $\pi f_1(1285)$ and $\eta f_1(1285)$ in Ref.~\cite{Jia:2026ewk}, the $K X(3872)$ in Ref.\cite{Song:2026spz}, the $\bar K^0$-deuteron in Ref.~\cite{Bayar:2026ted} and the $K^-$-deuteron in Ref.~\cite{Feijoo:2026eof}. 
More recently, it has been used to study the interaction of $\phi$ with $h_1(1380)$ \cite{Khemchandani:2026syv} which produces a state with the same quantum numbers, mass and width as the $X(2370)$ state, offering an alternative picture for that state different from the glueball nature claimed in Ref.~\cite{BESIII:2026mvn}. 
Interestingly, in all cases except for the $\pi f_1(1285)$ and $\eta f_1(1285)$, one obtains bound states with roughly $30-50 \mev$ binding, adding more states to the list of three-body systems, made solely from mesons. 

In the present work, we study the interaction of a kaon with the $T_{cc}(3875)$. 
The $T_{cc}$, discovered in Refs.~\cite{LHCb:2020bls,LHCb:2021vvq}, is also considered to be a molecular state of isospin $I=0$, stemming from the $D^0 D^{*+}, D^{*0} D^+$ interaction. 
The existence of such a bound system was anticipated in some theoretical works \cite{Li:2012ss,Liu:2019stu,Ding:2020dio}, and the analogous $D^* D^*$ system was also predicted to be bound in Ref.~\cite{Molina:2010tx}. 
The proximity of the $T_{cc}(3875)$ state to the $D D^*$ threshold makes it likely to be a molecular state of these components, as the general study of near-threshold structures of Ref.~\cite{Dong:2021rpi} would suggest. 
However, a detailed study of the $T_{cc}(3875)$ system \cite{Dai:2023kwv} indicates that the proximity of the system to the $D D^*$ threshold does not necessarily imply that the state is of molecular nature. 
It was shown in Ref.~\cite{Dai:2023kwv} that a compact nature is not excluded in principle, but one pays a big price for it, since the scattering length would be very small and the effective range extremely large, contradicting present values from Ref.~\cite{LHCb:2021vvq}. 
This allowed one to conclude in Ref.~\cite{Dai:2023kwv} that the non-molecular nature of this state is ruled out. 
The same conclusion was reached in Ref.~\cite{Dai:2023cyo}  from the study of the experimental spectra, allowing the contribution of a compact state, which was rejected by the data. 

Actually, the $T_{cc}$ state was explained within the molecular picture in Ref.~\cite{Feijoo:2021ppq}, predicting much smaller widths than originally reported in Ref.~\cite{LHCb:2020bls}, which were later corrected in Ref.~\cite{LHCb:2021vvq}. 
After that, the list of papers supporting the molecular picture of the $T_{cc}$ is large \cite{Ling:2021bir,Dong:2021bvy,Yan:2021wdl,Xin:2021wcr,Ren:2021dsi,Albaladejo:2021vln,Du:2021zzh,Ke:2021rxd,Agaev:2022ast,Padmanath:2022cvl,Albaladejo:2022sux,Cheng:2022qcm,Wang:2022jop,Sun:2024wxz,Gil-Dominguez:2024zmr}.

The study of the $K T_{cc}$ system is very promising for generating a new three-body mesonic state. 
Indeed, with the $D^* D$ nature of the $T_{cc}$ state, the $K$ interacts with the $D^*$ to produce the $D_{s1}(2460)$ state \cite{Kolomeitsev:2003ac,Gamermann:2007fi,Guo:2006rp,Gil-Dominguez:2023puj,Wu:2022wgn,Faessler:2007us}. 
On the other hand, the $K$ interacting with the $D$ produces the $D^*_{s0}(2317)$ \cite{Kolomeitsev:2003ac,Guo:2006fu,Xie:2010zza,Liu:2012zya,Gamermann:2006nm,Guo:2015dha,Faessler:2007gv}. 
Both states, $D^*_{s0}(2317)$ and $D_{s1}(2460)$, are bound by about $50 \mev$ with respect to their $D^* K$ and $D K$ components; hence, we expect the kaon to act as a glue to further bind the $D^*$ and $D$. 
This would be analogous to the case of a kaon interacting with a (unbound) $p p$ system, which leads to a bound state of $K^- pp$, predicted by many theoretical works (see detailed discussions in Ref.~\cite{Bayar:2026ted}) and confirmed experimentally in Refs.~\cite{J-PARCE15:2018zys,J-PARCE15:2020gbh}. 
Since the $K D, K D^*$ interactions are stronger than that of $\bar K N$, which produces the $\Lambda(1405)$ states, less bound than the $D^*_{s0}(2317)$ and $D_{s1}(2460)$, we anticipate a very strongly bound state of the  $K D^* D$ system. 
By looking at the quark content, one starts from a $c \bar d c \bar u$ ($D^{*+} D^0$) and goes to a $c c \bar u \bar s$ state ($K^0 D^{*+} D^0$) after a hypothetical annihilation of $d \bar d$, which makes the three-body bound state as exotic as the $T_{cc}$. 

Interestingly, the system that we propose has already been the subject of investigation from different perspectives. 
In Ref.~\cite{Ren:2024mjh} the system was studied from the same perspective as the present one using the FCA method, and the system was found bound with about the same binding as found here. 
In the present work, we use the FCA implementing elastic unitarity, which allows us to study the interaction close to threshold and evaluate the $K^0 T_{cc}$ correlation function. 
In Ref.~\cite{Ma:2017ery}, the $D D^* K$ system is studied with a very different framework and dynamics, using a delocalized pion bond with the pion shared by the three particles, and the system is also found bound with about the same binding energy. 
A different conclusion was reached in Ref.~\cite{Tan:2024omp}, where a quark model was used and a small binding energy of about $0.8 \mev$ was found for the system.
In Ref.~\cite{Zhang:2024yfj} the $DD^*K$ system was studied within lattice effective field theory, looking at the effect of three-body forces. 
Their results showed that without the three-body interaction, the $DD^*K$ system was bound by about $79 \mev$. 
By increasing the three-body repulsive interaction, the binding energy of $DD^*K$ gradually decreased, approaching
about $45 \mev$ in the limit of the infinitely strong three-body repulsive interaction. 
A study along these lines using the Gaussian expansion method \cite{Pan:2025xvq} corroborated these results and found that in the limit where the strength of the three-body potential becomes infinite, the $DD^*K$ bound state would break into a $D^* K$ two-body bound state, accompanied by a distant $D$ meson.  
In this respect it is mandatory to make some comments concerning the three-body forces. 

In a dynamical calculation, the strength of a three-body force is tied to the model used for the two-body forces. 
It is well known that the off-shell extrapolation of the two-body scattering amplitudes generates a contact three-body force. 
Different models leading to the same on-shell two-body amplitudes generate different three-body forces. 
Then the additional three-body force needed to describe a three-body system is tied to the two-body model used. 
There is no absolute three-body force. 
In this regard, it is worth recalling the findings of Refs.~\cite{MartinezTorres:2008gy,MartinezTorres:2007sr}, which showed that, when the standard chiral Lagrangians for meson-meson \cite{Gasser:1983yg} and meson-baryon \cite{Ecker:1994gg} interactions are used, the three-body force generated by the off-shell contributions of the two-body amplitudes in the Faddeev approach is exactly cancelled by the three-body force arising from the same Lagrangians.
Hence, our approach, where these Lagrangians are used to generate the on-shell two-body amplitudes needed in the FCA, without any further three-body force, makes implicit use of these findings, rendering the approach self-consistent (see a detailed discussion of this issue in Ref.~\cite{Encarnacion:2025lyf}).

We shall study the interaction of a $K^0$ with the $T_{cc}$ using the unitary FCA \cite{Ikeno:2025bsx,Agatao:2025ckp}. 
The formalism is developed in Section \ref{sec:form}; 
in Section \ref{sec:res} we show results for the scattering amplitude and the bound states obtained, the low energy scattering parameters and the correlation function, 
and in Section \ref{sec:con} we finish with some conclusions. 

\section{Formalism}\label{sec:form}
We follow closely the formalism of Ref.~\cite{Jia:2026dpl} for the study of the $K f_1(1285)$ interaction.
There the $K$ is interacting with a $K^* \bar{K}$ cluster in $I = 0$.
Here we take the external particle to be a $K^0$ to avoid dealing with Coulomb effects in the correlation function.
The cluster here is $D^* D$ in $I=0$, so the analogy is obvious: one replaces the $K K^*$ interaction by $K D^*$ and the $K \bar{K}$ interaction by $K D$. 
Although the analogy is clear from the formal point of view, the dynamics of the amplitudes is rather different, and so are the results.

\begin{figure}[b]
	\begin{center}
		\includegraphics[width=0.49\textwidth]{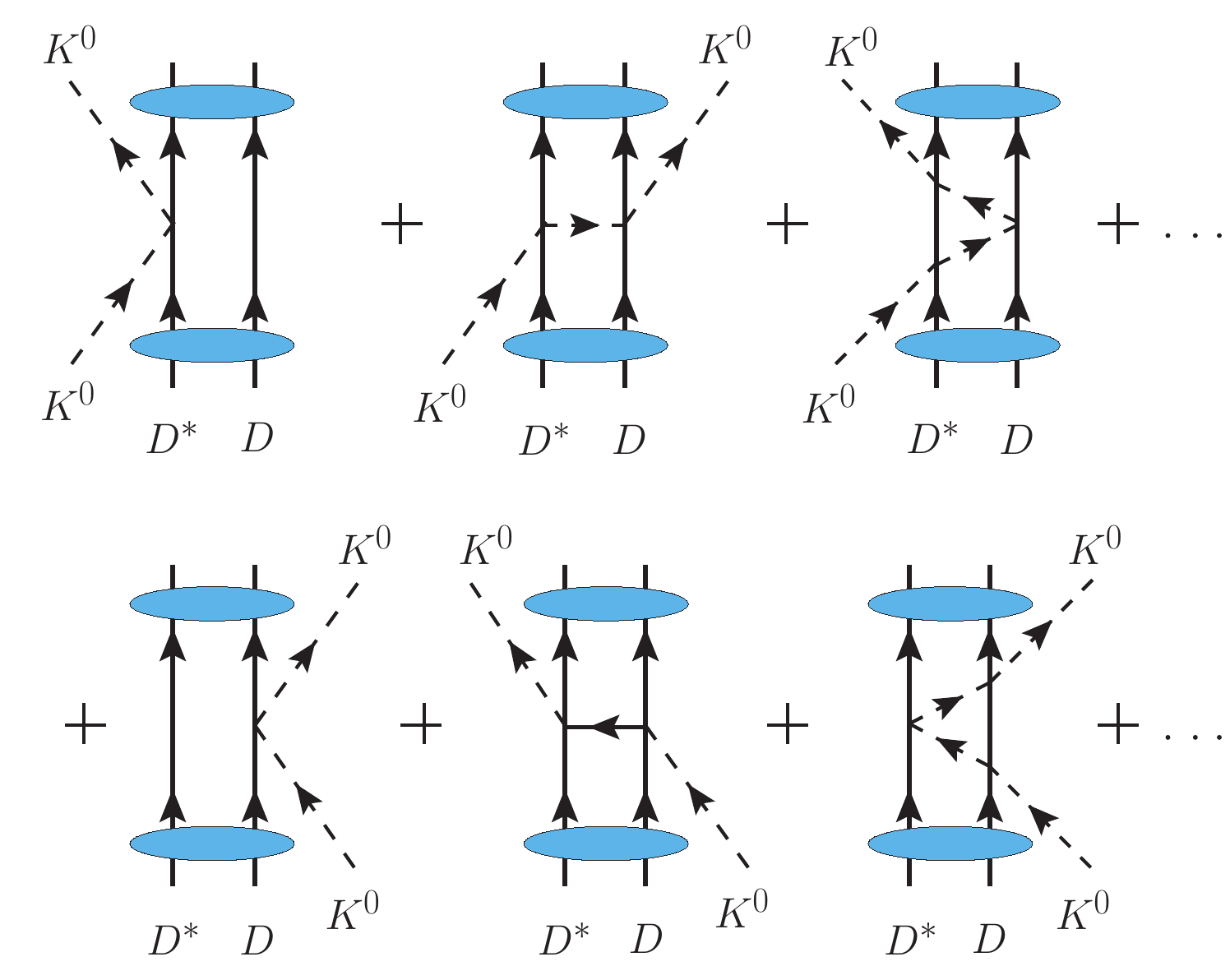}
	\end{center}
	\vspace{-0.5cm}
	\caption{Diagrams entering the ordinary FCA approach for $K^0$ interacting with the cluster of the $D^* D$.}
	\label{Fig1}
\end{figure}

We start by summing the diagrams of the standard FCA, depicted in Fig.~\ref{Fig1}.
Denoting the $KD^*$ and $KD$ amplitudes by $t_1$ and $t_2$, respectively, and considering that $K$ interacts with the $D^*D$ system in the $I=0$ channel, we have
\begin{equation}
\begin{aligned}
	t_1 &= \dfrac{3}{4}\;t_{KD^*}^{I=1}+\dfrac{1}{4}\;t_{KD^*}^{I=0},\\[2mm]
	t_2 &= \dfrac{3}{4}\;t_{KD}^{I=1}+\dfrac{1}{4}\;t_{KD}^{I=0},
	\label{eq:1}
\end{aligned}
\end{equation}
where the $t_{KD}^{I=0, 1}$ and $t_{KD^*}^{I=0, 1}$ amplitudes are calculated in Appendix \ref{sec:appA} for the cases without and with the $\pi^0-\eta$ mixing.

Next, we will write a final amplitude for the $K^0 T_{cc}$ system.
To adjust it to the field factors of these particles, we define
\begin{equation}\label{eq:2}
\tilde{t}_1 = \frac{M_c}{m_{D^*}} t_1,  \quad \tilde{t}_2 = \frac{M_c}{m_D} t_2, 
\end{equation}
where $M_c$ is the mass of the $T_{cc}$ state.

Then, one defines the partition functions $\tilde{T}_{ij}$ which sum all the diagrams of Fig.~\ref{Fig1} in which the $K^0$ interacts first with particle $i$ and finishes with particle $j$, and the matrix $\tilde{T}$ given by
\begin{equation}
	\tilde{T}= \begin{pmatrix} \tilde{T}_{11} & \tilde{T}_{12} \\[1mm] \tilde{T}_{21} & \tilde{T}_{22} \end{pmatrix},
	\label{eq:3}
\end{equation}
where
\begin{equation}
	\begin{split}
		& \tilde{T}_{11} = \dfrac{\tilde{t}_1}{1 - \tilde{t}_1 \,\tilde{t}_2 \,G_0^2}, ~~~~~~~~~~~~
		\tilde{T}_{22} = \dfrac{\tilde{t}_2}{1 - \tilde{t}_1\,\tilde{t}_2 \,G_0^2}, \\[2mm]
		& \tilde{T}_{12} = \tilde{T}_{21} = \dfrac{\tilde{t}_1 \,\tilde{t}_2 \, G_0}{1 - \tilde{t}_1 \,\tilde{t}_2  \,G_0^2}.
	\end{split}
	\label{eq:4}
\end{equation}
The function $G_0(\sqrt{s})$ in Eq.~\eqref{eq:4} is the $K^0$ propagator from particle 1 to particle 2 in the cluster,  modulated by the wave function of the cluster, given by
\begin{align}\label{eq:5}
	G_0(\sqrt{s}) &= \int \dfrac{\mathrm{d}^3 q}{(2\pi)^3} 
	\;\dfrac{F_c(q)}{\sqrt{s} - \omega_{K^0}(\vec q\,) - \omega_c(\vec q \,) + i \epsilon} \;\dfrac{1}{2 \,\omega_{K^0}(\vec q\,)} \nonumber \\[2mm]
	&\times \dfrac{1}{2 \,\omega_{c}(\vec q\,)}  \;\Theta\left(q_{\mathrm{max}}^{(1)} - q_1^*\right)  \Theta\left(q_{\mathrm{max}}^{(2)} - q_2^*\right),
\end{align}
with $\sqrt{s}$ the center-of-mass energy of the $K^0 T_{cc}$ system, 
and $\omega_{j}(\vec q\,) = \sqrt{m_{j}^2 + \vec{q}^{\;2}}$,
where $F_c(q)$ is the form factor of the cluster, given by
\begin{equation}\label{eq:6}
	\begin{aligned}
		F_c(q) &= \dfrac{F(q)}{N}, \\[2mm]
		F(q) &= \int\limits_{\substack{|\vec{p}\,| < q_{\mathrm{max}} \\ |\vec{p} - \vec{q}\,| < q_{\mathrm{max}}}} 
		\dfrac{\mathrm{d}^3 p}{(2\pi)^3} \;
		\dfrac{1}{M_c - \omega_{D^*}(\vec{p}\,) - \omega_{D}(\vec{p}\,)} \\
		&\quad \times \dfrac{1}{M_c - \omega_{D^*}(\vec{p} - \vec{q}\,) - 
			\omega_{D}(\vec{p} -\vec{q}\,)}, \\[2mm]
		N &= F(0) \\
		&= \int\limits_{|\vec{p}\,| < q_{\mathrm{max}}} \dfrac{\mathrm{d}^3 p}{(2\pi)^3} \,
		\left[ \dfrac{1}{M_c - \omega_{D^*}(\vec{p}\,) - \omega_{D}(\vec{p}\,)} \right]^2,
	\end{aligned}
\end{equation}
where $q_{\rm max}$ is the regulator of the $D^* D$ loops in the study of the $D^* D$ interaction of Ref.~\cite{Feijoo:2021ppq}, which we take here as $q_{\rm max} = 415 \mev$.
The $\Theta (\;)$ functions accompany the scattering matrices $t_i$ \cite{Gamermann:2009uq} and appear inside the $\vec q$ integration of $G_0(\sqrt{s})$, and $q_1^*$, $q_2^*$ are the $K^0$ momenta in the rest frame of $K^0D^*$ and $K^0D$ respectively, assuming the momenta transfered to be shared equally between the initial and final particle of the cluster in a collision~\cite{Carrasco:1991we,Boffi:1991nh} and is given by~\cite{Jia:2025obs}
\begin{equation}
	\vec{q}^{\,*}_i = \vec{q}\left( 1 - \frac{1}{2}\frac{m_{K^0}}{m_{K^0} + m_{i}} \right),
	\label{eq:qstar}
\end{equation}
where $m_1=m_{D^*}$, $m_2=m_{D}$. 
In Eq.~\eqref{eq:5} $q_{\mathrm{max}}^{(1)}$ and $q_{\mathrm{max}}^{(2)}$ are the regulators in the loops in the study of the $K^0 D^*$ and $K^0 D$ scattering amplitudes, respectively, which will be addressed in Appendix \ref{sec:appA}.

One must also specify the arguments $s_1$, $s_2$ of the scattering matrices $t_1$, $t_2$. They are given in terms of $s$,
\begin{equation}
	\begin{split}
		s_1 &= \left(p_{K^0} + p_{D^*}\right)^2= m_{K^0}^2 + \left(\xi \, m_{D^*}\right)^2 + 2\xi \, m_{D^*} \,q^0,\\[2mm]
		s_2 &= \left(p_{K^0} + p_{D}\right)^2= m_{K^0}^2 + \left(\xi \, m_{D}\right)^2 + 2\xi \, m_{D} \,q^0,
	\end{split}
	\label{eq:s1s2}
\end{equation}
where $q^0$ is the $K^0$ energy in the cluster rest frame
\begin{equation}\label{eq:9}
	q^0=\frac{s-m_{K^0}^2-M_{c}^2}{2\,M_{c}},
\end{equation}
and
\begin{equation}\label{eq:10}
	\xi=\frac{M_{c}}{m_{D^*} + m_{D}}.
\end{equation}
Equation~\eqref{eq:s1s2}, through the factor $\xi$, assumes that the binding of the $T_{cc}$ cluster is shared by the $D^*$ and $D$ particles proportionally to their masses. 
In the present case, with the small binding of the $T_{cc}$, this consideration has no practical effects.

\begin{figure}[t]
	\begin{center}
		\includegraphics[width=0.49\textwidth]{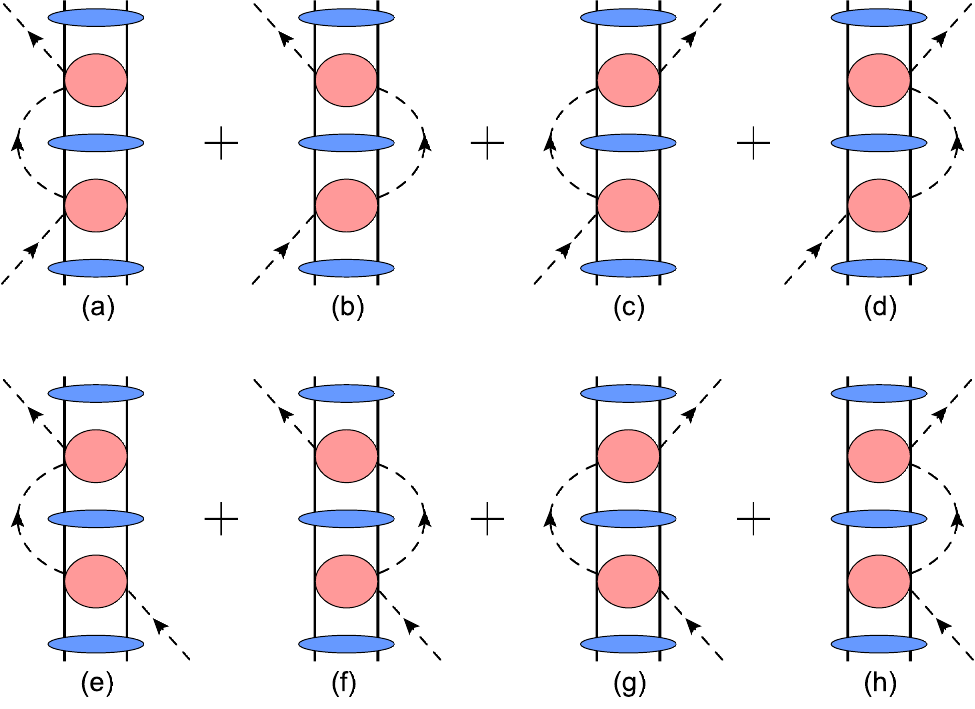}
	\end{center}
	\vspace{-0.5cm}
	\caption{Diagrams considering the elastic propagation of the $K^0$ and the cluster of $T_{cc}$ as a whole.}
	\label{Fig2}
\end{figure}

The unitarity of the $T$ matrix for $K^0 T_{cc}$ scattering is achieved by summing the diagrams of Fig.~\ref{Fig2}.
Analytically the diagrams are summed to the new partition functions $\tilde{T}'_{ij}$, gathered in the matrix $\tilde{T}'$
\begin{equation}
	\tilde{T}' = \begin{pmatrix} \tilde{T}_{11}' & \tilde{T}_{12}' \\[1mm]
		\tilde{T}_{21}' & \tilde{T}_{22}' \end{pmatrix}, \\
	\label{eq:11}
\end{equation}
which is obtained via
\begin{equation}\label{eq:12}
	\tilde{T}'=\left[ 1-\tilde{T}\,G_c \right]^{-1}\tilde{T},
\end{equation}
where
\begin{equation}\label{eq:13}
	G_c=\begin{pmatrix}
		G_c^{(1)} & 0 \\[1mm]
		0 & G_c^{(2)}
	\end{pmatrix},
\end{equation}
with
\begin{equation}\label{eq:14}
	\begin{aligned}[b]
		G_c^{(i)}(\sqrt{s}) &= \int \frac{\mathrm{d}^3 q}{(2\pi)^3} 
		\frac{\left[ F_c^{(i)}(\vec q \,) \right]^2}{\sqrt{s} - \omega_{K^0}(\vec q\,) - \omega_c(\vec q\,) + i\epsilon} \\[1.5mm]
		&\quad \times \dfrac{1}{2\omega_{K^0}(\vec q\,)} \dfrac{1}{2\omega_{c}(\vec q\,)}\,\Theta\left(q_{\mathrm{max}}^{(i)} - q_i^*\right),
	\end{aligned}
\end{equation}
with
\begin{equation}
	\begin{aligned}
		F_c^{(1)}(\vec q \,) &= F_c\left( \frac{m_{D}}{m_{D^*}+m_{D}} \, \vec q\right),\\[2mm]
		F_c^{(2)}(\vec q\,) &= F_c\left( \frac{m_{D^*}}{m_{D^*}+m_{D}} \, \vec q\right).
	\end{aligned}
	\label{eq:15}
\end{equation}

A simplified expression is obtained in Ref.~\cite{Agatao:2025ckp} for the total $K^0 T_{cc}$ scattering matrix
\begin{align}\label{eq:16}
	T_{K^0 T_{cc}}^{\mathrm{tot}}&=\sum_{i,j} \tilde{T}'_{ij}\nonumber\\[2mm]
	&=\frac{
		\tilde{t}_1+\tilde{t}_2+\left(2G_0-G_c^{(1)}-G_c^{(2)}\right) \tilde{t}_1 \, \tilde{t}_2}
	{1-G_c^{(1)} \tilde{t}_1-G_c^{(2)}\tilde{t}_2-\left(G_0^2-G_c^{(1)} G_c^{(2)}\right) \tilde{t}_1 \, \tilde{t}_2},
\end{align}
and an analytical proof of the unitarity of this matrix is provided in Ref.~\cite{Agatao:2025ckp}. 
Unitarity implies that the scattering amplitude $T_{K^0 T_{cc}}^{\mathrm{tot}}$ has a threshold behaviour given by
\begin{equation}\label{eq:fQM}
	\begin{aligned}[b]
		-8\pi\sqrt{s} \,\left(T_{K^0 T_{cc}}^{\mathrm{tot}}\right)^{-1} &= (f^{\mathrm{QM}})^{-1} \\[1mm]
		&\approx -\frac{1}{a} + \frac{1}{2}\,r_0 \,q_{\mathrm{cm}}^2 - i q_{\mathrm{cm}},
	\end{aligned}
\end{equation}
where $f^{\mathrm{QM}}$ is the ordinary Quantum Mechanics scattering matrix, and $q_{\mathrm{cm}}$ is the $K^0$ momentum in the $K^0 T_{cc}$ rest frame. 
In Eq.~\eqref{eq:fQM}, the scattering length $a$ and the effective range $r_0$ depend on the dynamics, but the linear term in $q_{\mathrm{cm}}$, $-iq_{\mathrm{cm}}$, is universal and is a consequence of the unitarity of the scattering matrix.

In terms of our amplitude $T_{K^0 T_{cc}}^{\mathrm{tot}}$, $a$ and $r_0$ are given by
\begin{equation}\label{eq:18}
	\begin{aligned}
	a &= \frac{T_{K^0 T_{cc}}^{\mathrm{tot}}}{8\pi\sqrt{s}}\Big|_{\mathrm{th}}, \\[2mm]
	r_0 &= \frac{1}{\mu} \left[ \frac{\partial}{\partial \sqrt{s}} \left( -8\pi\sqrt{s} \,\left(T_{K^0 T_{cc}}^{\mathrm{tot}}\right)^{-1} + iq_{\mathrm{cm}} \right) \right]_{\mathrm{th}}, 
	\end{aligned}
\end{equation}
with $\mu$ the $K^0T_{cc}$ reduced mass.

The correlation function for the $K^0 T_{cc}$ system is given by
\begin{equation}
	\begin{aligned}[b]
		C_{K^0 T_{cc}}(p) = & 1 + 4\pi \int_{0}^{\infty} \mathrm{d}r \, r^2\,  S_{12}(r) \\[1.5mm]
		& ~~~~~\times \left\{ \left|j_0(pr) + TG\right|^2 - j_0^2(pr) \right\},
	\end{aligned}
	\label{eq:19}
\end{equation}
where
\begin{equation}
	TG = T_{K^0 T_{cc}}^{\mathrm{tot}}\;\frac{1}{2} \left[ G_1(\sqrt{s},r) + G_2(\sqrt{s},r) \right],
	\label{eq:20}
\end{equation}
and $S_{12}(r)$ is the source function
\begin{equation}
	S_{12}(r)=\frac{1}{\left(
		4\pi R^2
		\right)^{3/2}}\;e^{-r^2/4R^2},
	\label{eq:21}
\end{equation}
with
\begin{equation}
	\begin{aligned}[b]
		G_i(\sqrt{s},r) &= \int \frac{\mathrm{d}^3 q}{(2\pi)^3} \;
		\frac{ j_0(qr)\; F_c^{(i)}(q) }{\sqrt{s} - \omega_{K^0}(\vec q\,) - \omega_c(\vec q\,) + i\epsilon} \\[1.5mm]
		&\quad \times \dfrac{1}{2\omega_{K^0}(\vec q\,)} \;\dfrac{1}{2\omega_{c}(\vec q\,)} \;\Theta\left(q_{\mathrm{max}}^{(i)} - q_i^*\right).
	\end{aligned}
	\label{eq:22}
\end{equation}

\section{Results}\label{sec:res}
\begin{figure}[b]
	\begin{center}
		\includegraphics[width=0.48\textwidth]{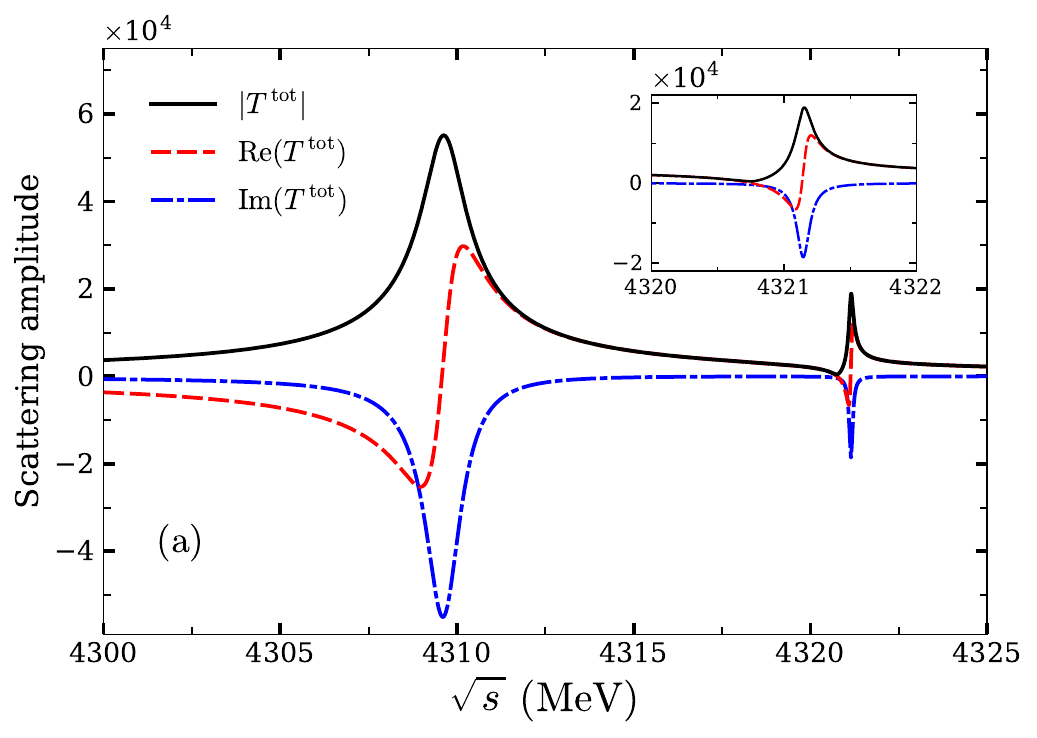}\\
		\includegraphics[width=0.48\textwidth]{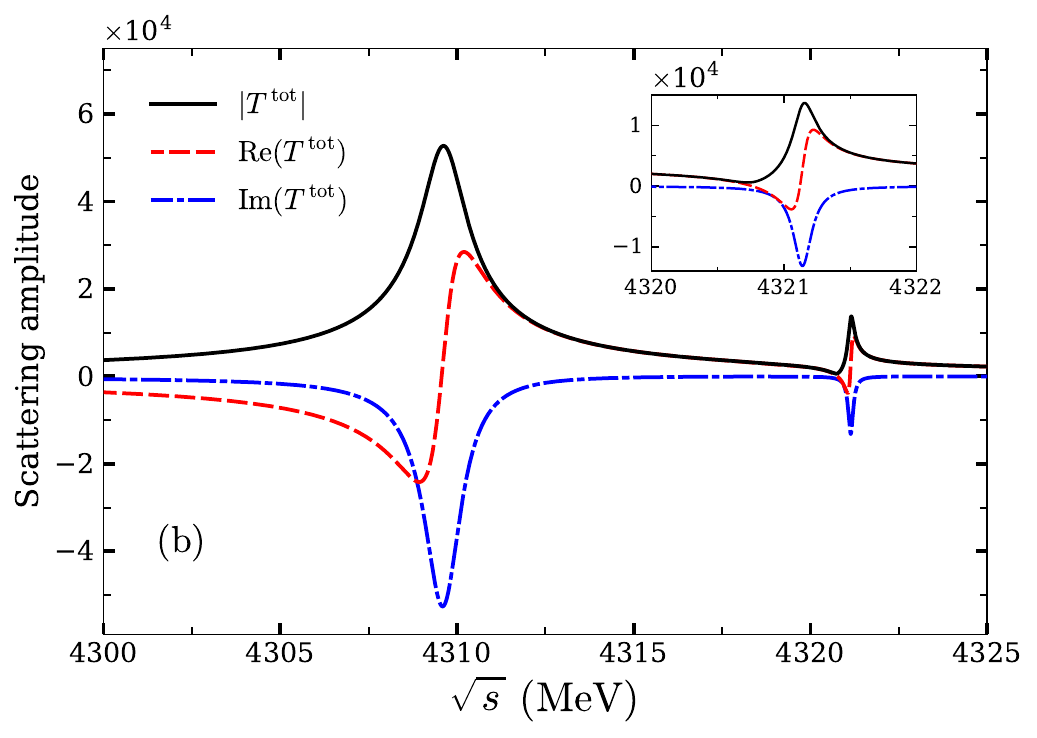}
	\end{center}
	\vspace{-0.5cm}
	\caption{The total three-body scattering amplitude $T_{K^0 T_{cc}}^{\mathrm{tot}}$ as a function of the three-body center-of-mass energy $\sqrt{s}$, with (a) and (b) corresponding to the cases without $\pi^0-\eta$ mixing and with $\pi^0-\eta$ mixing, respectively. The insets show magnified views of the resonant structure at higher energy.}
	\label{Fig:amplitude} 
\end{figure}

In Fig.~\ref{Fig:amplitude}, we show the result for the total amplitude $T_{K^0 T_{cc}}^{\mathrm{tot}}$ for the cases without $\pi^0-\eta$ mixing [Fig.~\ref{Fig:amplitude}(a)] and with $\pi^0-\eta$ mixing [Fig.~\ref{Fig:amplitude}(b)].
It is notable that two resonant structures exist in the amplitude: one around $4310\mev$ with a larger strength, and another around $4321\mev$ with a much smaller strength.
These two resonant structures have the feature that $\Re T_{K^0 T_{cc}}^{\mathrm{tot}}$ changes from negative to positive going through zero at the peak of $|\Im \, T_{K^0 T_{cc}}^{\mathrm{tot}}|$.
From the width of $|\Im \, T_{K^0 T_{cc}}^{\mathrm{tot}}|$, we extract a width of the lower energy state of about $1.21\mev$ from Fig.~\ref{Fig:amplitude}(a) or $1.26\mev$ from Fig.~\ref{Fig:amplitude}(b), and the maximum of $|\Im \, T_{K^0 T_{cc}}^{\mathrm{tot}}|$ occurs at $4309.62\mev$, which we take as the mass of this state.
For the higher energy state, we extract the mass of $4321.2\mev$ and a width of about $0.123\mev$ from Fig.~\ref{Fig:amplitude}(a) or $0.180\mev$ from Fig.~\ref{Fig:amplitude}(b).
For the case with $\pi^0-\eta$ mixing, we can also look for poles in the complex plane of $T^{\rm tot}_{K^0 T_{cc}}$ by making $\sqrt{s}$ complex, without modifying any of the $G$ functions except for making $\sqrt{s}$ complex. 
This defines our Riemann sheet.
We obtain a pole at $\sqrt{s} = 4309.62+ i\, 0.63 \mev$ for the first resonant structure, and another pole at $4321.2+i \, 0.08 \mev$ for the second resonant structure, in remarkable agreement with the values of the mass and width extracted from $|\Im \, T_{K^0 T_{cc}}^{\mathrm{tot}}|$. 
The resonant structures obtained correspond to two bound states of the $KT_{cc}$ system, with bindings of about $63\mev$ and $51\mev$ with respect to the $K^0 T_{cc}$ threshold.
We should note that the consideration of the $\pi^0-\eta$ mixing in the $t_1, t_2$ amplitudes, described in Appendix \ref{sec:appA}, does not change the mass of the state, and the width is only increased by about $5\%$ for the first peak.

\begin{figure}[t]
	\begin{center}
		\includegraphics[width=0.48\textwidth]{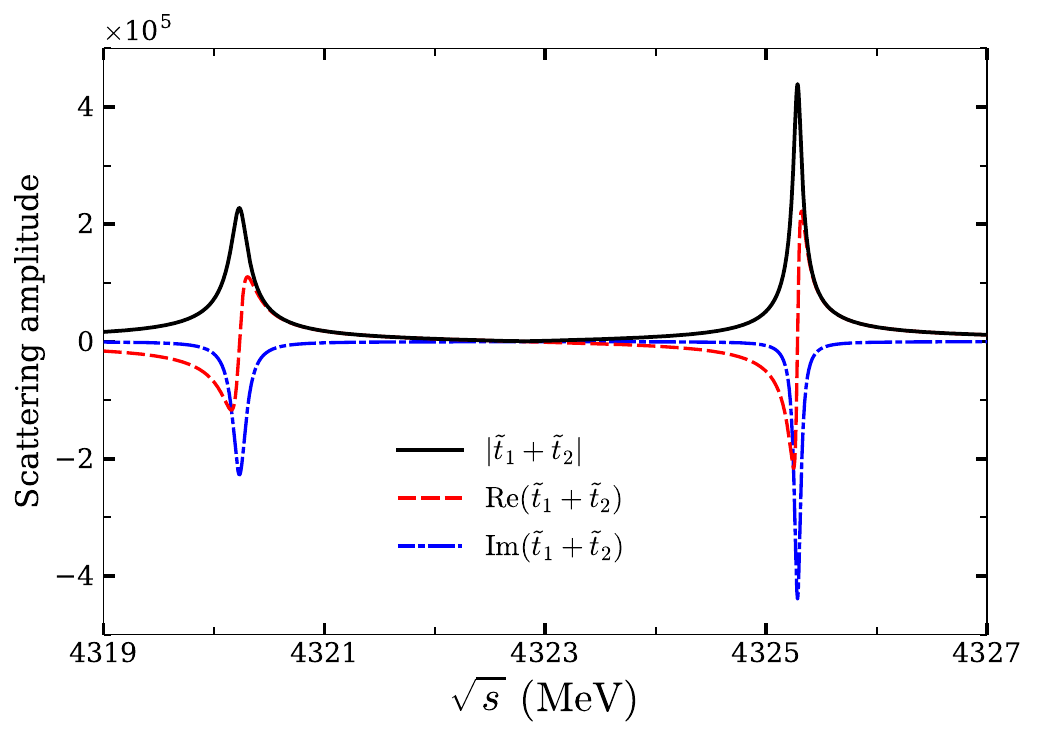}
	\end{center}
	\vspace{-0.5cm}
	\caption{The total amplitude for $KT_{cc}$ in the impulse approximation $T_{K^0 T_{cc}}^{\rm tot}=\tilde{t}_1 + \tilde{t}_2$, with $\pi^0-\eta$ mixing.}
	\label{Fig:amplitude_Impulse}
\end{figure}

Compared with the results of Refs.~\cite{Ikeno:2025bsx,Jia:2026dpl,Encarnacion:2026zas,Jia:2026iqo}, where the same unitary FCA formalism was applied to the $n \bar D \bar K$, $K K^* \bar K$, $p K^* \bar K$ and $KDK$ three-body systems and only one bound state was found in each case, one may wonder why two bound states are found for the $K D^* D$ interaction in the present work. 
The reason is tied to the fact that the $K D$ and $K D^*$ interactions are both attractive and strong enough to generate the $D^*_{s0}(2317)$ and $D_{s1}(2460)$ molecular states.
To further understand it, we show in Fig.~\ref{Fig:amplitude_Impulse} the results for the total $KT_{cc}$ amplitude in the impulse approximation $T_{K^0 T_{cc}}^{\rm tot}=\tilde{t}_1 + \tilde{t}_2$, where only single scatterings (the first two diagrams of Fig.~\ref{Fig1} to the left) are considered.
Two neat peaks are observed in the amplitude, one at around $4320\mev$ and the other at around $4325\mev$.
The strength of the second peak is much bigger than that of the first one, in contrast to the case in Fig.~\ref{Fig:amplitude}.
Compared with the results of Fig.~\ref{Fig:amplitude_Impulse}, the two peaks in Fig.~\ref{Fig:amplitude} are shifted to lower energies by several MeV and become wider, which is a consequence of the multiple scatterings taken into account in the unitary FCA formalism.
The multiple scatterings also reduce the strength of the two peaks, with the second peak being reduced more significantly.

\begin{figure}[t]
	\begin{center}
		\includegraphics[width=0.49\textwidth]{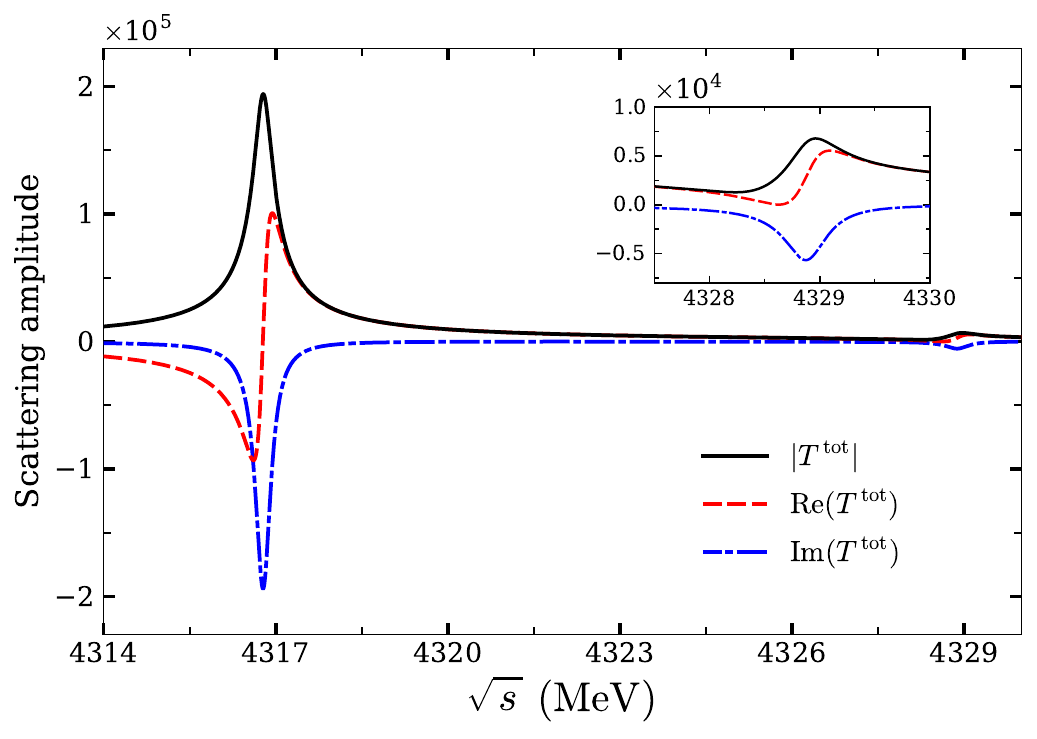}
	\end{center}
	\vspace{-0.5cm}
	\caption{$T_{K^0 T_{cc}}^{\mathrm{tot}}$ in the conventional (non-unitary) FCA formalism, with $\pi^0-\eta$ mixing. The inset provides a magnified view of the resonant structure at higher energy.}
	\label{Fig:T-nonunitary}
\end{figure}

It is also interesting to show the results that we obtain for $T_{K^0 T_{cc}}^{\mathrm{tot}}$ in the conventional (non-unitary) FCA formalism. 
This is accomplished in our case by setting $G_{c}^{(1)} = G_{c}^{(2)} = 0$ in Eq.~\eqref{eq:16}.
We show in Fig.~\ref{Fig:T-nonunitary} the result for $T_{K^0 T_{cc}}^{\mathrm{tot}}$ in this case.
We observe several features: The $T_{K^0 T_{cc}}^{\mathrm{tot}}$ has again two resonant structures similar to that of the unitary amplitude; 
The positions of the two peaks are slightly different to those of the unitary amplitude, with the first peak at $4316.8\mev$ and the second peak at $4328.8\mev$;
The strength of the first peak is bigger than that of the unitary amplitude, by about a factor of 3.5, but the strength of the second peak is very similar. 
This corroborates a known fact that the standard FCA is a fair tool to predict bound states of three-body system. 
The energy of the first peak at $4316.8 \mev$ should be compared with the one of Ref.~\cite{Ren:2024mjh} of $4309 \mev$, which was calculated with no width.  
Our results with the unitary FCA with binding of $63 \mev$ are also in line with those of Refs.~\cite{Ma:2017ery,Zhang:2024yfj,Pan:2025xvq} within the uncertainties mentioned there from their three-body force. The novelty is that we have observed a second $KT_{cc}$ bound state with a mass of $4321.2 \mev$ and a width of about $0.2\mev$, which is not found in the works mentioned above.

The scattering length and effective range obtained are
\begin{equation}
\begin{aligned}
	a &= 0.50-i 0.24~ \text{fm},\\[2mm]
	r_0 &= 2.44-i 2.05~ \text{fm}.
	\label{eq:aandr0}
\end{aligned}
\end{equation}
The effect of including the $\pi^0-\eta$ mixing only changes these results below $5\%$.

\begin{figure}[t]
	\begin{center}
		\includegraphics[width=0.4\textwidth]{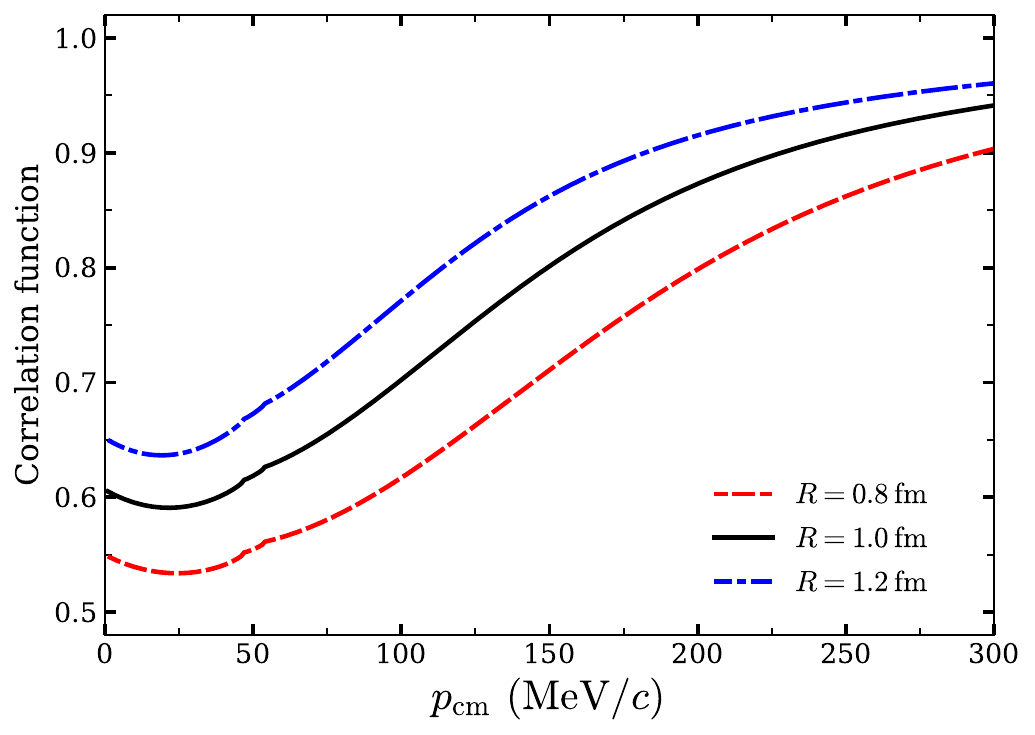}\\
		\includegraphics[width=0.4\textwidth]{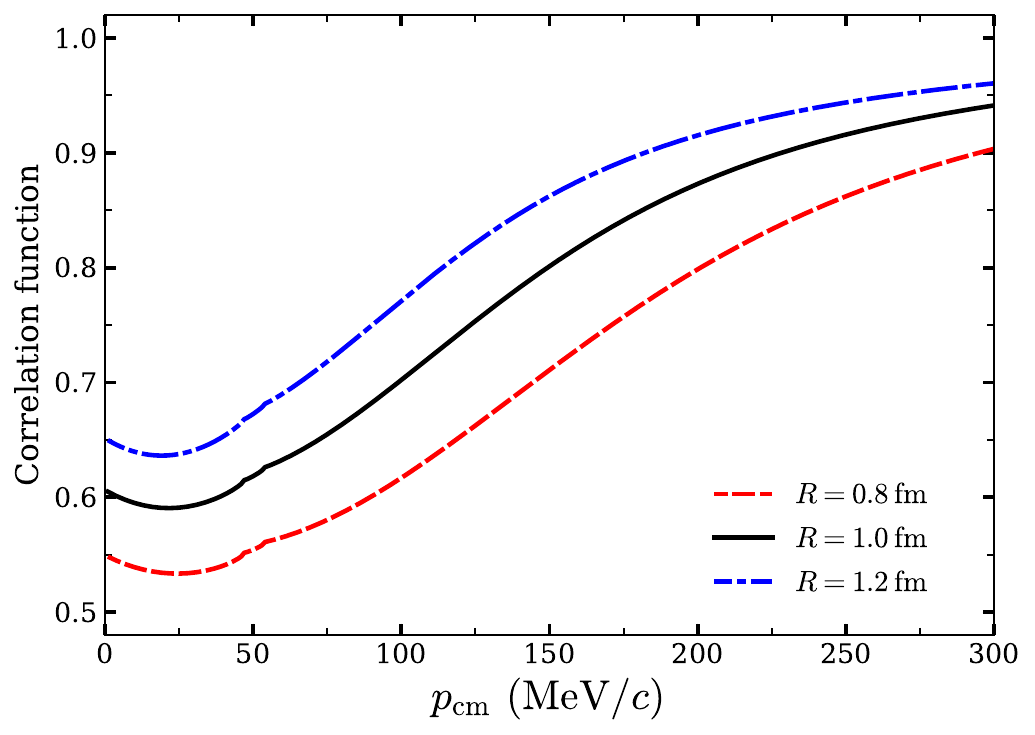}
	\end{center}
	\vspace{-0.5cm}
	\caption{Correlation functions for $K^0T_{cc}$ with different $R$, with the upper and lower panels corresponding to the cases without $\pi^0-\eta$ mixing and with $\pi^0-\eta$ mixing, respectively. The first kink in the correlation function at $p_{cm} \simeq  47\mev/c$ is caused by the $D^{*+}K^0$ threshold of the two-body interaction, while the second kink at $p_{cm} \simeq  54\mev/c$ is due to the $D^+K^0$ threshold, from $t_1$ and $t_2$ amplitudes, respectively, via Eq.~\eqref{eq:s1s2}.}
	\label{Fig:CF}
\end{figure}

Finally, we show in Fig.~\ref{Fig:CF} the correlation function, which has a shape similar to those obtained in previous cases mentioned above where there was also a bound state of the three-body system. 
There is a substantial deviation from unity at low relative momentum $p_{cm}$, and the correlation function increases gradually to reach values close to unity around $300 \mev/c$. 
According to the classification in Ref.~\cite{Liu:2023uly}, this behaviour corresponds to the case of a strong attractive interaction capable of producing a bound state.

\begin{figure}[t]
	\begin{center}
		\includegraphics[width=0.48\textwidth]{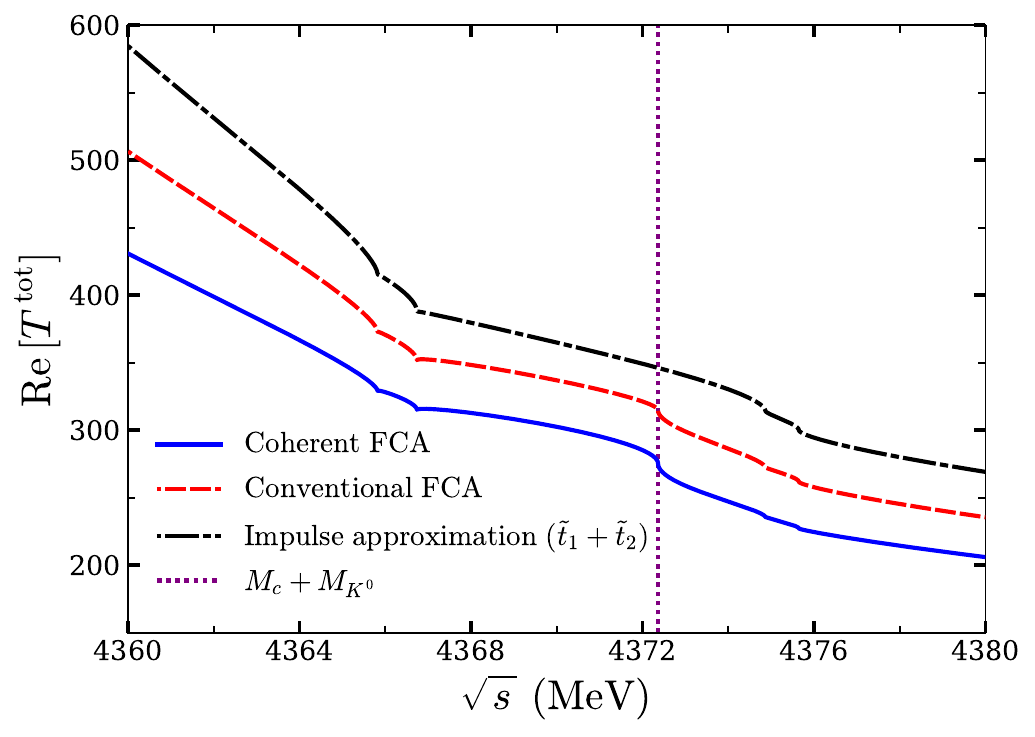}
	\end{center}
	\vspace{-0.5cm}
	\caption{Comparison of $\Re T_{K^0 T_{cc}}^{\mathrm{tot}}$ in different formalisms.}
	\label{Fig:Compar_ReT}
\end{figure}

\begin{figure}[t]
	\begin{center}
		\includegraphics[width=0.48\textwidth]{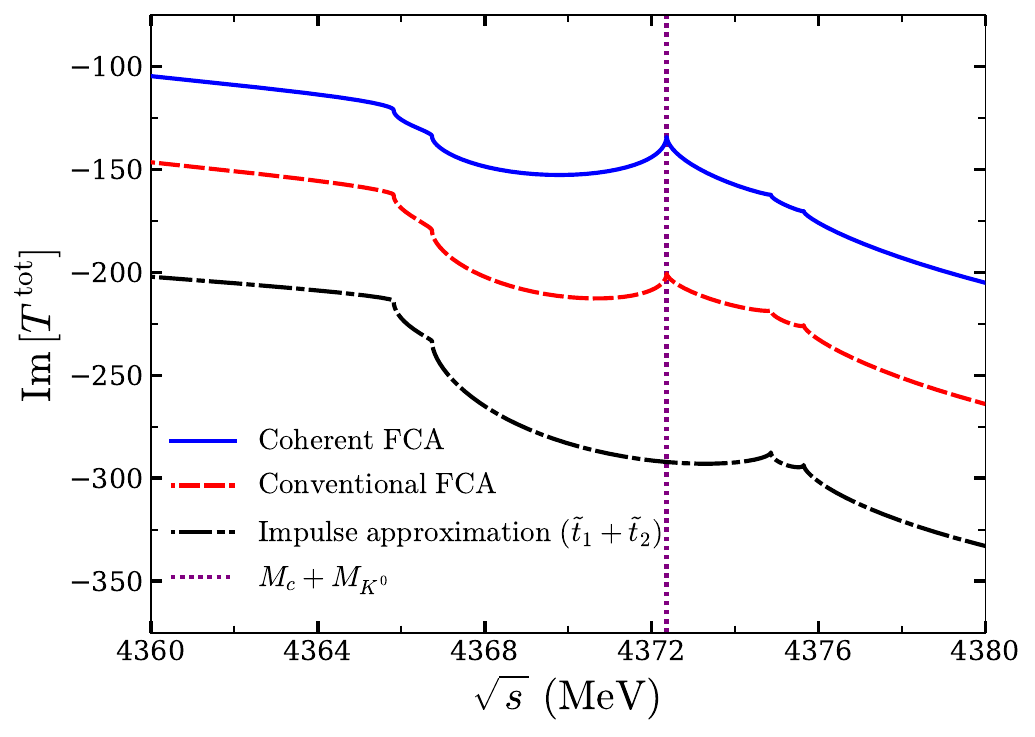}
	\end{center}
	\vspace{-0.5cm}
	\caption{Comparison of $\Im T_{K^0 T_{cc}}^{\mathrm{tot}}$ in different formalisms.}
	\label{Fig:Compar_ImT}
\end{figure}

Coming back to the work of Ref.~\cite{Ren:2024mjh}, the main differences with respect to the unitary $T_{K^0 T_{cc}}^{\mathrm{tot}}$ matrix appear, as expected, close to threshold.
In Figs.~\ref{Fig:Compar_ReT} and \ref{Fig:Compar_ImT} we show the real and imaginary parts of the amplitude $T_{K^0 T_{cc}}^{\mathrm{tot}}$ in three cases: impulse approximation $T = \tilde{t}_1 + \tilde{t}_2$, the conventional FCA [with $G_{c}^{(1)} = G_{c}^{(2)} = 0$ in Eq.~\eqref{eq:16}] and the unitary or coherent FCA [Eq.~\eqref{eq:16}].
We observe that $\Re T_{K^0 T_{cc}}^{\mathrm{tot}}$ decreases by going from the impulse approximation to the coherent FCA. 
The difference between the conventional and coherent FCA at threshold are of the order of $10\%$. 
The changes are bigger in $\Im T_{K^0 T_{cc}}^{\mathrm{tot}}$ shown in Fig.~\ref{Fig:Compar_ImT}. 
The difference between the conventional and coherent FCA are of the order of $30\%$. 
In Fig.~\ref{Fig:Compar_ImT}, we observe some small kinks in the amplitude, starting from left to right they correspond to the $K^+ D^0$, $K^+ D^{*0}$, $K^0 T_{cc}$, $K^0 D^{*+}$ and $K^0 D^+$ thresholds, respectively.

We can thus see that the scattering amplitudes around threshold are appreciably changed by using the coherent FCA approach versus the conventional one, and, as already mentioned, the effective range $r_0$ requires a unitary approach to be obtained.
Without imposing unitarity in the amplitude, the linear term in $q_{cm}$, which is incorrectly given by the conventional FCA amplitude, dominates the change of $T_{K^0 T_{cc}}^{\mathrm{tot}}$ at low energies above threshold, making the determination of $\frac{1}{2} r_0 q_{cm}^2$ term unreliable.

\section{Conclusions} \label{sec:con}
We have addressed the problem of the interaction of a kaon with the $T_{cc}(3875)$ in search of a new exotic system made of three mesons. 
For this, we adopt the molecular picture of the $T_{cc}(3875)$ as a bound state of $D^* D$ in isospin $I=0$, and allow the kaon to interact with this two-body system. 
One can anticipate that the system should be bound. 
Indeed, the $K$ interacting with the $D^*$ produces the $D_{s1}(2460)$ resonance, while the $DK$ interaction gives rise to the $D^*_{s0}(2317)$, both of which are bound by about $50 \mev$ with respect to the threshold mass of their components. 
The kaon, thus, acts as a glue that brings the $D^* D$ closer together and produces two $KD^*D$ bound states with binding energies of about $63 \mev$ and $51\mev$ respectively, with respect to the $KT_{cc}(3875)$ threshold. 
The system bears resemblance to the $K^- pp$ bound state, which originates from the interaction of kaon with a $pp$ system. 
Even if the $pp$ system is unbound, the kaon provides the glue to bring the two protons together. 
The experimental observation of this latter three-body state in two different laboratories, J-PARC and LEPS, gives extra credibility to the prediction of the state made in the present work. 
The state predicted is exotic, both in the sense of corresponding to a three-body system, and also from the quark content containing a minimum $c c \bar{u} \bar{s}$ which cannot annihilate into a system with fewer quarks.

We have also investigated the scattering amplitude near threshold and highlighted the importance of recent developments in implementing unitarity in the Fixed Center Approximation (FCA), 
which is employed in the present work. 
Using this approach, we have determined the scattering length and effective range for $K^0 T_{cc}$ scattering. 
We have also calculated the $K^0 T_{cc}$ correlation function and found features similar to those observed in other studies of the scattering of external particles off resonances that are dynamically generated through the interaction of their constituent particles. 
Recent advances by the ALICE Collaboration in the study of correlation functions involving stable particles and resonances, as well as those involving $D$ mesons, 
together with the growing interest of the LHCb Collaboration in investigating correlation functions for more complex systems, 
offer promising prospects for the experimental investigation of the phenomena predicted in the present work in the near future.
 
\section*{Acknowledgments}
This work is partly supported by the Natural Science Foundation of Guangxi province under Grant No. 2023JJA110076,
and by the National Natural Science Foundation of China (NSFC) under Grants No. 12575081 and No. 12365019.
This work is also partly supported by the National Key R{\&}D Program of China (Grant No. 2024YFE0105200).
This work is also partly supported by the Spanish Ministerio de Economia y Competitividad (MINECO) and European FEDER funds under Contracts No. FIS2017-84038-C2-1-PB, PID2020-112777GB-I00, and by Generalitat Valenciana under contract PROMETEO/2020/023. 
This project has received funding from the European Union Horizon 2020 research and innovation program under the program H2020-INFRAIA-2018-1, grant agreement No. 824093 of the STRONG-2020 project.

\appendix
\section{$t_1,\,t_2$ amplitude}\label{sec:appA}
\begin{figure}[b]
	\begin{center}
		\includegraphics[width=0.45\textwidth]{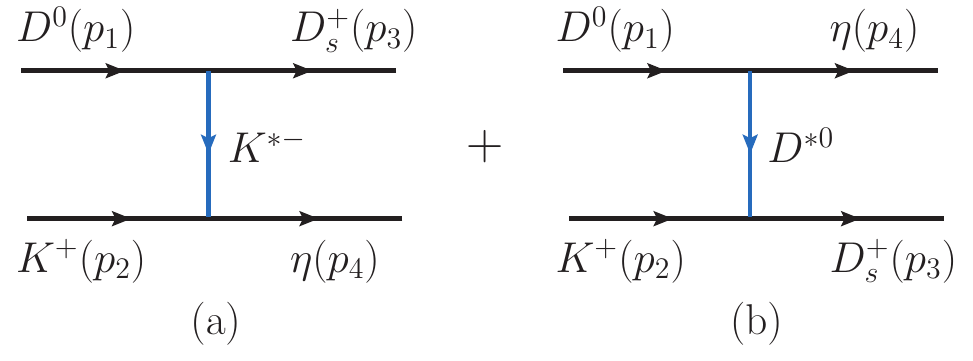}
	\end{center}
	\vspace{-0.5cm}
	\caption{The vector-meson exchange mechanism for $i\to j$ transitions: direct diagram (a) and crossed diagram (b).}
	\label{FigA1}
\end{figure}

The $KD$ amplitudes in $I=0$ and $I=1$ are obtained in Ref.~\cite{Su:2025aiz} using the coupled channel $D^0 K^+ \,(1)$, $D^+ K^0\, (2)$, $D^+_s \eta \,(3)$, $D^+_s \pi^0\,(4)$.
The $i\to j$ transitions proceed via the vector-meson exchange mechanism, as shown in Fig.~\ref{FigA1}.

With the phase convention $(K^+,\,K^0)$, $(\bar K^0,\,-K^-)$, $(D^+,\, -D^0)$, $(\bar D^0,\,D^-)$, the isospin states are
\begin{equation}
	\begin{aligned}
		| DK,\, I=0 \rangle &= \frac{1}{\sqrt{2}}\left( D^+ K^0 + D^0 K^+\right),\\[1mm]
		| DK,\, I=1, I_3=0 \rangle &= \frac{1}{\sqrt{2}} \left( D^+ K^0 - K^0 K^+ \right).
	\end{aligned}
	\label{eq:A1}
\end{equation}
Then the $KD$ isospin amplitudes are 
	\begin{align}\label{eq:A2}
		&t_{KD}^{I=0}=\langle 
		DK, \,I=0 \left|\, t \,\right|DK,\,I=0 \rangle \nonumber\\[1mm]
		=&\frac{1}{2}\langle D^+ K^0 + D^0 K^+ | t| D^+ K^0 + D^0 K^+  \rangle \nonumber\\[1mm]
		=&\frac{1}{2}(
		t_{D^+K^0,D^+K^0}+t_{D^0K^+,D^0K^+}+2t_{D^+K^0,D^0K^+}),
	\end{align}
	\begin{align}\label{eq:A2p}
		&t_{KD}^{I=1}=\langle 
		DK, \,I=1 \left|\, t \,\right|DK,\,I=1 \rangle \nonumber\\[1mm]
		=&\frac{1}{2}\langle D^+ K^0 - D^0 K^+ | t| D^+ K^0 - D^0 K^+  \rangle \nonumber\\[1mm]
		=&\frac{1}{2}(
		t_{D^+K^0,D^+K^0}+t_{D^0K^+,D^0K^+}-2t_{D^+K^0,D^0K^+}),
	\end{align}
where the two-body amplitudes $t_{ij}$ are obtained from the Bethe-Salpeter equation in coupled channels, 
\begin{equation}
	T = \left[1-VG\right]^{-1}V,
\end{equation}
with $V_{ij}$ the transition potential given in Ref.~\cite{Su:2025aiz}, and $G$ the meson-meson loop function regularized with a cutoff. 
As in Ref.~\cite{Su:2025aiz}, we take a cutoff $q^{(2)}_\text{max} = 687 \mev$, which reproduces the mass of the $D^*_{s0}(2317)$.

\begin{figure}[t]
	\begin{center}
		\includegraphics[width=0.23\textwidth]{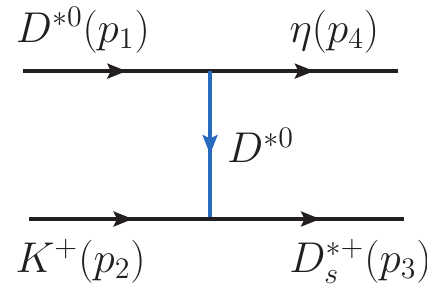}
	\end{center}
	\vspace{-0.5cm}
	\caption{Diagram equivalent to Fig.~\ref{FigA1}(b), with $D$ replaced by $D^*$.}
	\label{FigA2}
\end{figure}

The $KD^*$ amplitudes are obtained in the same way. 
The coupled channels are now $D^{*0} K^+ \,(1)$, $D^{*+}K^0\, (2)$, $D^{*+}_s \eta \,(3)$, $D^{*+}_s \pi^0\,(4)$. 
Formally the transition potentials are the same, changing the mass of $D$ to $D^*$. 
There is one subtle difference. 
In Ref.~\cite{Su:2025aiz}, there are terms that go as $C'_{ij} \cdot g(p_i)$. 
They correspond to crossed diagrams like the one in Fig.~\ref{FigA1}(b).
In the $D^*K$ interaction, this corresponds to the diagram of Fig.~\ref{FigA2}.
However, while in Fig.~\ref{FigA1}(b) the upper vertex is of the type vector-pseudoscalar-pseudoscalar, the one in Fig.~\ref{FigA2} is vector-vector-pseudoscalar, which is anomalous and smaller than the other vertex. 
In addition, both diagrams involve the exchange of a heavy vector and are suppressed with respect to the other terms exchanging light vectors. 
We then ignore the $g(p_i)$ terms of the potentials for $D^*K$ interaction. 
We take a cutoff $q^{(1)}_\text{max}$ the same as for the $DK$ interaction, $q^{(2)}_\text{max} = 687 \mev$.

We also implement the $\pi^0-\eta$ mixing, as done in Ref.~\cite{Su:2025aiz}. One has
\begin{equation}
	\begin{aligned}
		&\tilde \pi^0 = \pi^0 \cos{\epsilon} + \eta \sin{\epsilon},\\[1mm]
		&\tilde \eta = -\pi^0 \sin{\epsilon} + \eta \cos{\epsilon},
	\end{aligned}
	\label{eq:A3}
\end{equation}
where $\tilde\pi^0$ and $\tilde\eta$ are the physical states and $\pi^0,\,\eta$ those of SU(3), and $\sin \epsilon = 0.01$. 
Considering the mixing of Eq.~\eqref{eq:A3} , the transition potential $V_{ij}$ are changed to $\tilde V_{ij}$, given by
\begin{align*}\label{eq:Vijp}
	\tilde{V}_{13}&= V_{13} \cos \epsilon -V_{14} \sin \epsilon,\nonumber\\[1mm]
	\tilde{V}_{14}&= V_{14} \cos \epsilon +V_{13} \sin \epsilon,\nonumber\\[1mm]
	\tilde{V}_{23}&= V_{23} \cos \epsilon -V_{24} \sin \epsilon,\nonumber\\[1mm]
	\tilde{V}_{24}&= V_{24} \cos \epsilon +V_{23} \sin \epsilon,\\
	\tilde{V}_{33}&= V_{33} \cos^2 \epsilon +V_{44} \sin^2 \epsilon -2 V_{34} \sin \epsilon \cos \epsilon, \nonumber\\[1mm]
    \tilde{V}_{34}&= V_{34} (\cos^2 \epsilon -\sin^2 \epsilon) +(V_{33} -V_{44}) \sin \epsilon \cos \epsilon,\nonumber\\[1mm]
	\tilde{V}_{44}&=V_{44} \cos^2 \epsilon +V_{33} \sin^2 \epsilon+2V_{34}\sin \epsilon \cos \epsilon.\nonumber
\end{align*}
Note that $V_{11}, V_{12}, V_{22}$ are not changed.

\bibliographystyle{a}
\bibliography{refs}
\end{document}